\documentclass[%
 aip,
 jmp,%
 amsmath,amssymb,
 reprint,%
]{revtex4-2}

\usepackage{graphicx}
\usepackage{dcolumn}
\usepackage{bm}

\begin{document}

\preprint{AIP/123-QED}

\title{Lagrangian Finite-Time Fluctuation Relation in isotropic turbulence}

\author{
Hanxun Yao$^{1}$, Tamer A.\,Zaki$^{1}$ and Charles Meneveau$^{1}$}

\address{$^{1}$Department of Mechanical Engineering \& IDIES, Johns Hopkins University}

\date{\today}

\begin{abstract}
The entropy generation rate in turbulence can be defined using the energy cascade rate as described in the scale-integrated Kolmogorov-Hill equation at a specified length scale. The fluctuation relation (FR) from non-equilibrium thermodynamics, which predicts exponential behaviour of the ratio of probability densities for positive and negative entropy production rates, was confirmed in prior work \citep{yao2023entropy}, but under certain limiting assumptions. We here examine the applicability of FR to isotropic turbulence under less stringent assumptions by analyzing entropy generation rates averaged over intervals ranging from one to several eddy turnover times. Based on time-resolved data at a Taylor-scale based Reynolds number $Re_\lambda = 433$, we find that the FR 
is valid in the sense that very close to exponential behaviour of probability ratios of positive and negative entropy generation (forward and inverse cascade of energy) is observed.  Interestingly, finite-time averaging yields FR-consistent results only within a Lagrangian framework, along fluid  trajectories using  filtered convective velocities. In contrast, the FR does not hold with  time-averaging at fixed (Eulerian) positions. Results provide evidence that the definition of entropy generation based on the scale-integrated Kolmogorov-Hill equation describes  turbulent cascade processes that exhibit properties predicted by non-equilibrium thermodynamics.
\end{abstract}

\maketitle

\section{Introduction}
\label{sec:introduction}

Entropy is a  macroscopic property of matter that increases and reaches a maximum when a system approaches thermodynamic equilibrium, corresponding to the most probable state subject to constraints. This fundamental principle underlies the second law of thermodynamics and establishes a bridge between reversible microscopic dynamics and the emergence of irreversible macroscopic behaviour. In non-equilibrium systems when either the number of discrete microscopic constituents (e.g. particles) is not large, the time-scale of microscopic interactions is not much faster compared to macroscopic elapsed time, or both, even macroscopic quantities such as entropy fluctuate over time. For such systems the rate of entropy generation and the process of reaching equilibrium are of particular interest. It can be negative during short periods of time, in apparent violation of the second law of thermodynamics, but quite allowable for systems away from thermodynamic equilibrium. From the viewpoint of dynamical systems theory, energy dissipation is associated with the contraction of phase-space volume (physically interpreted as entropy production \citep{gallavotti2020ensembles}). In the non-equilibrium thermodynamic framework, one of the fundamental and testable laws is the Fluctuation Relation (FR) \citep{evans1993probability, gallavotti1995dynamical, searles2000generalized, marconi2008fluctuation, seifert2012stochastic}. The FR describes the exponential relationship between the ratio of probability densities of observing a ``forward positive dissipative'' event and the corresponding ``negative dissipation reverse'' event, proportional to the local entropy generation rate. It is intimately related to the idea of exponential time-evolution of volumes in phase-space, and concomitant definitions of Lyapunov and Finite Time Lyapunov exponents (FTLE) of nonlinear dynamical systems.

For a trajectory bundle in phase space, an initial phase-space volume $V(0)$ typically evolves over time as
\begin{equation}
V(\tau) = V(0) \, \exp(-\Lambda^\tau \,\tau),
\end{equation} where $\Lambda^\tau$ represents the finite-time average of the phase-space contraction rate over the time interval $\tau$ \citep{gallavotti1995dynamical, evans2002fluctuation, ruelle1999smooth, ott2002chaos}.
Also, let $\Psi_{\cal S}(t)$ denote the total (net) macroscopic entropy generation rate of system ${\cal S}$ at time $t$. Following the system during time interval $\tau$, the total entropy generated during $\tau$ is given by $\Delta S_{\cal S} = \Psi^{\tau}_{\cal S} \, \tau$, where $\Psi^{\tau}_{\cal S}$ is the averaged finite time entropy generation rate
\begin{equation} \Psi^{\tau}_{\cal S}(t) = \frac{1}{\tau} \int_{t}^{t+\tau} \Psi_{\cal S}(t') \,dt'.
   \label{averagePsi}
\end{equation}
Since the phase-space contraction rate is interpreted as net entropy generation rate \citep{gallavotti1995dynamical, evans2002fluctuation, ruelle1999smooth, ott2002chaos}, we can write 
\begin{equation}
V(t+\tau) = V(t) \, \exp(-\Psi_{\cal S}^\tau \,\tau)
\end{equation}
for a bundle of states that are characterized by a given value of $\Psi_{\cal S}^\tau$ over a time $\tau$ and occupy a volume $V(t)$ at $t$.

For a microscopically reversible system, time-reversed trajectories result in an inverse transformation of the volume:
\begin{equation}
V(t) = V(t+\tau) \exp(\Psi_{\cal S}^\tau \, \tau).
\end{equation}
These forward and inverse processes can be characterized by the entropy generation rate $\Psi_{\cal S}^\tau$ and its time-reversed counterpart $-\Psi_{\cal S}^\tau$. This symmetry forms the basis for the FR, which compares the probabilities of forward and time-reversed trajectory bundles. With the further assumption that the probability density $P(\Psi_{\cal S}^\tau)$ to observe a bundle of initial conditions that are characterized by a particular given value of $\Psi_{\cal S}^\tau$ is proportional to the bundle volume in phase-space $V(t)$  \citep{yao2023entropy}, the FR then arises as a statistical statement about the ratio of phase-space volumes:
\begin{equation}
\frac{P(\Psi_{\cal S}^\tau)}{P(-\Psi_{\cal S}^\tau)} = \exp(\Psi_{\cal S}^\tau \, \tau).
\label{time-aver}
\end{equation}
The ratio of probabilities of observing forward (positive entropy generation) and reverse (negative entropy generation) events follows an exponential law involving the entropy production rate, a result encapsulated by the FR. This framework provides a statistical foundation for quantifying time-reversal asymmetry in non-equilibrium systems. A comprehensive review of fluctuation theorems and their physical interpretations is given by \cite{seifert2012stochastic}. 

The FR implies that forward, dissipative events are exponentially more likely than their time-reversed, entropy-reducing counterparts. In general, the rate of entropy generation ($\Psi_{\cal S}^\tau$, an inverse time-scale) is determined by the time-scales associated with microscopic interaction processes (e.g. particle collisions), while the system elapsed time $t$ is taken normally to be macroscopic time intervals. Hence, for systems with large time-scale separation between micro and macroscopic dynamics, usually one has $\Psi_{\cal S}^\tau\, \tau >>1 $, and the probability of observing violations of the second law ($P(\Psi_{\cal S}^\tau <0)$) becomes exponentially small. For example, consider a particle system in kinetic theory of gases with collision time-scales on the order of $10^{-9}$ seconds ($\Psi_{\cal S}^\tau \sim 10^{9}$ 1/s) observed over macroscopic time-scales of microseconds $\tau \sim 10^{-6}$. The   probability of violation of the second law would be on the order of one in $\sim\exp(1000)$, i.e. $1:10^{434}$ --- far more than the total number of particles in the universe. Hence for systems with large time-scale separations, the second law holds.

However, interesting trends appear when the time-scale separation becomes less extreme.   Physically, the FR predicts the probability of observing rare violations of the second law of thermodynamics over finite time intervals, implying occasional occurrences of negative entropy generation (indicative of inverse processes) over shorter time scales, comparable to those of the microscopic processes.  The FR laid the foundation for understanding how dynamical systems that exhibit microscopic reversibility while remaining statistically irreversible can lead to ``inverse processes''. The first reported observation of a transient violation of the second law of thermodynamics in a fluid mechanical system was made by \cite{evans1993probability}. In their seminal work on a 2D shear flow studied via molecular dynamics, they introduced a new definition of the natural invariant measure for trajectory segments. Through this formulation, they demonstrated that violations of the second law can occur over finite timescales, and that the probability of such violations decays exponentially with time and vanishes as time approaches large times compared to molecular time-scales, in accordance with the FR.

Fluid turbulence is considered to be a system far from equilibrium, in the sense of lack of time-scale separation. The cascade of energy which on average is from large to small scales proceeds among  scales of motion that are not very dissimilar: by the time the large-scales evolve and transfer some of their energy to small scales, the small-scales will have evolved only somewhat faster.  Recent studies have explored the intersection between turbulence and non-equilibrium thermodynamics. For instance, it has been shown that heat flux data for solar turbulent convection satisfies the FR \citep{viavattene2020testing}. The FR has also been tested in 3D turbulence \citep{alexakis2023fluctuation} considering temporal fluctuations of large-scale energetics in a closed flow (periodic box turbulence). \cite{nickelsen2013probing},
\cite{reinke2018universal} and \cite{fuchs2020small} studied a 1D Fokker–Planck equation model for the cascade process, expressed in terms of one-dimensional velocity increments, defining entropy changes along individual energy cascade trajectories. Based on the Fokker–Planck equation formalism, they successfully tested the various versions of the FR based on hot-wire single component turbulence data for probability density functions of longitudinal velocity increments.  

As mentioned above, the FR is expected to hold for systems with microscopic time reversibility, allowing to conclude that $\Psi_{\cal S}^\tau \to -\Psi_{\cal S}^\tau$ if time is run backwards. In turbulence, 
the dynamical evolution of eddies in the inertial range is thought to be unaffected (directly) by viscosity. The  truncated Euler equations are time-reversible at the equation level. And yet, turbulence demonstrates irreversible behaviour when these scales are coupled to additional modes and 
description levels that include the small, viscous scales (Navier-Stokes). This feature has been the subject of extensive research, including the works of \cite{she1993constrained},  \cite{carati2001modelling},  \cite{cichowlas2005effective},  \cite{domaradzki2007analysis}, \cite{eyink2009localness}, and \cite{cardesa2015temporal,cardesa2017turbulent}. 
\cite{vela2021entropy} examined the time-reversibility of inertial-range dynamics and related these dynamics to physical-space flow structures aiming to explain the observed asymmetry between forward (positive) and inverse (negative) cascade events. 

Since the FR could provide quantitative predictions of inverse (negative) cascade events, there is interest in a definition of entropy generation rate (and thus of entropy changes over finite times) that is directly and locally measurable from data in 3D turbulent flows, in a manner similar to which one can define terms such as turbulent kinetic energy, etc. Specifically, one seeks a local, physically motivated definition  which does not depend on specific energy cascade reduced model systems, or is limited to describing only the fluctuations of global energetics of entire systems. Such a definition was provided by \cite{yao2023entropy} who introduced a definition of the entropy generation rate $\Psi_{\ell}^\tau({\bf x},t)$ for a system consisting of a sphere of diameter $\ell$ (in the inertial range), whose ``microscopic constituents'' are thought to be the smaller-scale eddies (of scale smaller than $\ell$) inside the sphere. The definition of $\Psi_{\ell}({\bf x},t)$ (without any time-averaging) is based on a generalized Kolmogorov–Hill equation (also known as the Kármán–Howarth–Monin–Hill equation \citep{portela2017turbulence, yao2023role}). This formulation incorporates key local quantities, all defined at a prescribed inertial-range length scale $\ell$: the energy cascade rate, the viscous dissipation rate, as well as the so-called ``temperature of turbulence'', taken to be proportional to the turbulent energy of smaller scales. Using this framework,
direct numerical simulation (DNS) data could be used to measure $\Psi_{\ell}({\bf x},t)$ across the flow. The data confirmed, to very good approximation,  the applicability of the FR to turbulence \citep{yao2023entropy} and its conditional form based on local viscous dissipation  \citep{yao2024forward}.

However, these prior analyses were limited to analysis of fixed time snapshots, so that no time-averaging over a time-scale $\tau$ could be undertaken. Instead, instantaneous values of $\Psi_{\ell}({\bf x},t)$ were interpreted as representative of the averages over the eddy turning time corresponding to the eddies of scale $\ell$. As any inertial-range quantity in turbulence, it was conjectured that $\Psi_{\ell}({\bf x},t)$ changes over time-scales on the order of the eddy-turnover time $\tau_\ell$. Hence, $\tau$ was chosen to be equal to the constant eddy-turnover time $\tau_\ell$,  and 
$\Psi_{\ell}({\bf x},t) \, \tau_\ell \approx\Psi_{\ell}^\tau({\bf x},t) \, \tau$ was assumed. The goal of the present study is to   test the original, more fundamental, finite time-averaged FR over a range of time intervals $\tau$ using time-resolved data. Specifically, we investigate both an Eulerian perspective, in which time averaging is performed at fixed spatial locations, and a more physically justified Lagrangian perspective, in which averaging is performed along trajectories defined by spatially filtered velocity fields at scale $\ell$. 

In \S \ref{sec:define} we review the definition of local entropy generation rate for 3D turbulent flow as introduced in \cite{yao2023entropy}, and introduce a slight revision in the definition of the turbulence ``temperature'' (by including a free parameter, accounting for the partitioning of kinetic energy across the three velocity components but an unknown number of degrees of freedom). In \S \ref{sec:constanttime} we repeat the analysis of \cite{yao2023entropy} for single snapshot data but for isotropic turbulence at lower Reynolds number, for which the time-evolution is available. Results confirm the validity of the results of 
\cite{yao2023entropy} also for this lower Reynolds number dataset, which is then used in the next section \ref{sec:finitetime}. There, we expand the analysis by following spheres in time in a Lagrangian sense, and evaluating finite time averaged entropy generation rates for various time intervals $\tau$. We describe results for both Eulerian and Lagrangian time evolutions and compare results. Conclusions are presented in \S \ref{sec:conclusion}.

\section{Definition of local entropy generation in 3D turbulence}
\label{sec:define}

The definition proposed in \cite{yao2023entropy} is based on the exact evolution equation for the energy contained in velocity increments at particular scales. The Kolmogorov-Hill (KH) equation, often also called the Karman-Howarth-Monin-Hill (KHMH) equation,  was derived by \cite{hill2001equations, hill2002exact}. It   describes the evolution of the  square of local velocity increments $(\delta u_i^2)$ at a specific physical position ${\bf x}$, scale ${\bf r}$ and time $t$, under the effects of viscous dissipation, viscous transport, advection, and pressure work. The instantaneous equation (omitting forcing terms for simplicity) reads,
\begin{equation}
\frac{\partial \delta u_i^2}{\partial t} + u^*_{j}\frac{\partial \delta u_i^2}{\partial x_j}  = 
-\frac{\partial \delta u_j\delta u_i^2}{\partial r_j}-\frac{8}{\rho}\frac{\partial p^*\delta u _i}{\partial r_i} 
+\nu \frac{1}{2} \frac{\partial^2 \delta u _i  \delta u _i}{\partial x_j \partial x_j
}+ 
2\nu \frac{\partial^2 \delta u _i \delta u _i}{\partial r_j \partial r_j}
-
4\epsilon^*,
\label{ins_KHMH_noint}
\end{equation}
where $\delta u_i= \delta u_i({\bf x},{\bf r},t) = u_i^+ - u_i^-$ represents the vector of velocity increments. The superscripts $+$ and $-$ refer to two points in the physical domain at ${\bf x} + {\bf r}/2$ and ${\bf x} - {\bf r}/2$, which are separated by the vector $r_i = x_i^+ - x_i^-$. The middle point between these two locations is $x_i = (x_i^+ + x_i^-)/2$, representing `locality' of the equation. The superscript $*$ indicates the average value between these two points, e.g., the two-point averaged dissipation $\epsilon^* = (\epsilon^+ + \epsilon^-)/2$, where $\epsilon^{\pm}$ is the ``pseudo-dissipation" given by $\epsilon = \nu \left(\partial u_i / \partial x_j \right)^2$ and $\nu$ represents the kinematic fluid viscosity. Next, we integrate equation $\ref{ins_KHMH_noint}$ over a sphere in ${\bf r}$-space up to a diameter equal to $\ell$, divide the equation by the volume of the sphere $V_\ell = \frac{4}{3} \pi \left(\frac{\ell}{2}\right)^3$ and by a factor of 4. These steps lead to  the scale-integrated form of the KH equation,

\begin{equation}
  \frac{\widetilde{d} k_\ell}{d t}   =  \Phi_\ell  - \epsilon_\ell + P_\ell - D_\ell -  \frac{\partial q_j}{\partial x_j},
 \label{KHMH_local2}
\end{equation}
where the material derivative is defined as $\tilde{d}/dt = \partial/\partial t + \tilde{u}_j\partial/\partial x_j$ with filtered advection velocity, $ 
 \tilde{u}_j \equiv 
 \frac{1}{V_\ell}\int_{V_{\ell}}   u_j^* \, d^3{\bf r}_s$. 
 The integration involves a radial vector ${\bf r}_s={\bf r}/2$ ranging between 0 and radius $\ell/2$ and over all angles of the sphere.  The transported kinetic energy variable $k_\ell$ is the local kinetic energy associated with all scales smaller than $\ell$, defined as
\begin{equation}
 k_\ell({\bf x},t) \equiv 
 \frac{1}{2 \,V_\ell}\int\limits_{V_{\ell}}  \frac{1}{2} \delta u_i^2({\bf x},{\bf r},t) \, d^3{\bf r}_s.
 \label{kinetic}
\end{equation}

The first term on the right of equation \ref{KHMH_local2} is interpreted as the local energy cascade rate \citep{yao2024comparing, yao2025analysis} at scale $\ell$ at position ${\bf x}$, since it originates from a divergence in scale-space, leading to a surface integral over the outer surface of the sphere of diameter $\ell$: 
\begin{equation}
\Phi_\ell({\bf x},t)  \equiv -\frac{3}{4\,\ell}\frac{1}{S_\ell}\oint\limits_{S_{\ell}} \delta u _i^2\,\delta u _j\,  \hat{r}_j dS. 
\label{cascade}
\end{equation}where $\hat{r}_j$ is the unit vector in the direction of $r$. The second term in Eq. \ref{KHMH_local2} is the local volume averaged rate of dissipation envisioned in Kolmogorov's refined similarity hypothesis \citep{kolmogorov1962refinement, yao2024forward},

\begin{equation}
\epsilon_\ell({\bf x},t) \equiv \frac{1}{V_\ell}\int\limits_{V_{\ell}}
 \epsilon^*({\bf x},{\bf r},t) d^3{\bf r}_s.
\end{equation}
Eq. \ref{KHMH_local2} also includes  
\begin{equation}
P_\ell  \equiv -\frac{6}{\ell}\frac{1}{S_\ell}\oint\limits_{S_{\ell}} \frac{1}{\rho} \, p^* \, \delta u _j \, \hat{r}_j \,dS,
\end{equation} which is the   pressure-velocity correlation, representing surface-averaged pressure work at scale $\ell$ (it is positive if the work is done on the system inside the volume $V_\ell$). The equation also contains viscous diffusion term $D_\ell$ which can be assumed to be small in the inertial range but becomes relevant when $\ell$ approaches the viscous range. Finally, the equation also contains position-divergence of the spatial flux of small-scale kinetic energy defined as,
\begin{equation}
q_j = 
 \frac{1}{V_\ell}\int\limits_{V_{\ell}} \frac{1}{2} \left(\delta u _i^2 \delta u^*_j \right) \, d^3{\bf r}_s,
\end{equation}where $\delta u^*_j \equiv u^*_j - \tilde{u}_j$. 

Equation \ref{KHMH_local2} represents the ``first law of thermodynamics'' for our  system of interest. As argued in \cite{yao2023entropy} in the context of turbulence, the thermodynamic system is analogous to a molecular system, where a group of eddies contained within a sphere of diameter $\ell$   are the analogue to a group of ``particles". Equation \ref{KHMH_local2} describes the energy evolution within this sphere, accounting for several key processes: energy exchange across length scale $\ell$ at a rate of $\Phi_\ell$, energy transport in physical space through a spatial flux $q_j$, pressure work exerted on the sphere’s periphery at a rate of $P_\ell$, and energy dissipation into molecular degrees of freedom at a rate of $\epsilon_\ell$.

To introduce the notion of entropy generation, \cite{yao2023entropy} applied the standard Gibbs equation
\begin{equation}
T \, ds = de + p \, dv,
\end{equation}where \( T \) is the temperature, \( e \) is the internal energy, and \( p \, dv \) represents the work done by pressure on the system. This equation is applied to the system of eddies within a sphere of size \( \ell \). To continue the analogy, \cite{yao2023entropy} made the following associations: \( p \, dv \sim -P_\ell \) and \( e \sim k_\ell \). The specific entropy of the system, the particular sphere of size $\ell$, was denoted as \( s_\ell \). By dividing by a time-increment \( dt \) and substituting the Lagrangian rate of change of kinetic energy from the energy conservation equation (Eq. \ref{KHMH_local2}),  an expression for the entropy generation rate was obtained:

\begin{equation}
   \frac{\widetilde{d} s_\ell}{dt} = \frac{1}{T} \left(\Phi_\ell - \epsilon_\ell - \frac{\partial q_j}{\partial x_j}\right).
 \label{KHMH_gibbs2}
\end{equation}
In this analysis, we are primarily interested in the entropy generation associated with the cascade rate $\Phi_\ell$ of kinetic energy across different scales  and not in the spatial fluxes of small-scale kinetic energy $q_j$. The last term is hence omitted from the analysis. Moreover,  the heat exchange with the ``thermal reservoir" at a rate of \( \epsilon_\ell \), occurring at temperature \( T \), results in a corresponding increase in the entropy of the reservoir at a rate given by \(  {\widetilde{d} s_{\rm res}}/{dt} =  {\epsilon_\ell}/{T} \). Consequently, the generation rate of the total entropy, defined as \( s_{\rm tot} = s_\ell + s_{\rm res} \), evolves according to:

\begin{equation}
   \frac{\widetilde{d} s_{\rm tot}}{dt} = \frac{\Phi_\ell}{T}.
 \label{entropytot}
\end{equation}
Additionally, following \cite{castaing1996temperature}, \cite{yao2023entropy} invoked the association
\( T \sim k_\ell \), taking as as possible definition of ``temperature'' the kinetic energy contained in the microscopic degrees of freedom. We here refine the definition of temperature of turbulence by recalling the relationship between internal (kinetic) energy and temperature for classical systems $e= \frac{1}{2}  n  k_B T$, where $k_B$ is the Boltzmann constant and $n$ the number of degrees of freedom ($3N$ for $N$ particles translating in 3 directions). Since for turbulence, neither the number of particles nor their degrees of freedom are defined or known, and the unit of ``temperature'' is (so far) arbitrary, we simply define the temperature $T$ as proportional to the small-scale translational (kinetic) energy $k_\ell$,
\begin{equation}
   T = \frac{2}{\gamma_t} \, k_\ell,
 \label{tempdef}
\end{equation}
where $\gamma_t$ will be determined from data, and represents a sort of ``turbulence Boltzmann constant times number of degrees of freedom'' (analogous to $nk_B$), and whose units are chosen as dimensionless so that ``T'' has units of kinetic energy. 

Using these definitions, it is natural to define the entropy generation rate according to
\begin{equation}
   \Psi_\ell({\bf x},t) = \frac{\gamma_t}{2}\, \frac{\Phi_\ell({\bf x},t)}{k_\ell({\bf x},t)}, 
 \label{entropygen}
\end{equation}
where $\Phi_\ell({\bf x},t)$ is defined, and can be measured from turbulent velocity data, according to Eq. \ref{cascade} and $k_\ell$ from Eq. \ref{kinetic}. 
The dimension of \( \Psi_\ell \) is inverse  time, analogous to the classical definition of entropy generation, i.e., heat transfer rate divided by temperature. This  definition of an  entropy generation rate is local in physical space (it describes a property of a fluid volume of scale $\ell$ located at position ${\bf x}$ and time $t$), and involves classical variables of turbulence theory (the angle-integrated third-order structure function, and the scale-integrated second order structure function). It makes no assumptions involving any cascade model process, Fourier modal decomposition, or even flow isotropy, or homogeneity. In this work we  will focus on homogeneous isotropic turbulence and  investigate the degree to which $\Psi_\ell({\bf x},t)$ adheres to the predictions from the FR, using both instantaneous and finite-time-averaged values of \( \Psi_\ell \), in the subsequent sections.

\section{Fluctuation relation analysis from  single-time data}
\label{sec:constanttime}
As summarized in \S \ref{sec:introduction}, prior work examined instantaneous snapshots from DNS of isotropic turbulence at a relatively high Reynolds number \citep{yao2023entropy}. There, the FR was tested under the assumption that 
$ \Psi_\ell \tau_\ell \approx \Psi_\ell^\tau \tau$, if $\tau_\ell = \langle\epsilon\rangle^{-1/3} \ell^{2/3}$ is used as characteristic eddy-turnover time scale of turbulent eddies at scale $\ell$. The main objective of this paper is to test the FR without having to make such an assumption, namely by using time-resolved DNS data. We have such data available but at a lower Reynolds number of $Re_\lambda \sim 433$. 

Before performing the time-dependent analysis, in this section we first confirm that the single-time observations made in \citep{yao2023entropy} also apply to this lower Reynolds number dataset.  For this purpose, we evaluate $\Psi_\ell$ from its definition at a single time (Eq. \ref{entropygen}). In order to be consistent with the approach used in \cite{yao2023entropy}, which used
as definition of $\Psi_\ell=\Phi_\ell/k_\ell$ instead of Eq. \ref{entropygen},  for the analysis in this section we set $\gamma_t=2$.  If the FR holds in the analysis of single-time instantaneous snapshots, plots of $\ln [P(\Phi_\ell/k_\ell)/P(-\Phi_\ell/k_\ell)]$ versus $(\Phi_\ell/k_\ell) \tau_\ell = (\Phi_\ell/k_\ell) \, \langle\epsilon\rangle^{-1/3} \ell^{2/3}$ should produce a straight line, i.e.

\begin{equation}
   \ln \left[\frac{P(\Phi_\ell/k_\ell)}{P(-\Phi_\ell/k_\ell)}\right] \propto  \frac{\Phi_\ell}{k_\ell},
 \label{FRrelation1}
\end{equation} 
with a slope of order unity (about 1.13 was found in \cite{yao2023entropy}). 

In the present study, we utilize direct numerical simulations of forced isotropic turbulence from the publicly available Johns Hopkins Turbulence Database (JHTDB; \cite{yao2025analysis}). 
As in \cite{yao2023entropy}, the analysis is conducted at three length scales within the inertial range: $\ell = 30\eta$, $45\eta$, and $60\eta$, where $\eta$ denotes the Kolmogorov length scale computed based on the overall rate of dissipation of the flow. To compute the surface averages required for evaluating $\Phi_\ell$, the outer surface of diameter $\ell$ is discretized into 500 approximately uniformly distributed point pairs (denoted as $+$ and $-$ points) on a sphere. The velocity components used to calculate $\delta u_i$ are obtained via the  JHTDB function `getData' using an 8th-order Lagrangian spatial interpolation scheme.
The quantity $k_{\ell}$ is similarly computed by a discretized volume integration by integrating over five concentric spherical shells. The accuracy of this surface and volume integration approach has been verified through convergence tests by increasing the number of discretization points \citep{yao2024comparing}. The quantities $k_\ell$, $\Phi_\ell$, and the rate $ \Phi_\ell / k_\ell$ (proportional to the entropy generation rate) are computed at one million spheres centered on random points, sampled across 2000 spatial locations within the three-dimensional domain $[0,2\pi]^3$, over 500 statistically stationary time snapshots during the available 5 large-scale turn-over time interval available in the database. Although $\Phi_\ell/k_\ell$ is computed across multiple time snapshots for statistical convergence of the PDFs, the present method does not involve any temporal averaging. Given the statistical stationarity in time and homogeneity in space characteristic of isotropic turbulence, each sample can be regarded as an independent realization of the local, instantaneous flow state. As a result, the combined set of spatio-temporal samples constitutes a valid statistical ensemble for constructing the PDF of $\Phi_\ell/k_\ell$ and and for testing the constant-time FR. 

\begin{figure} [h!]
 \centering
  \includegraphics[scale=0.35]{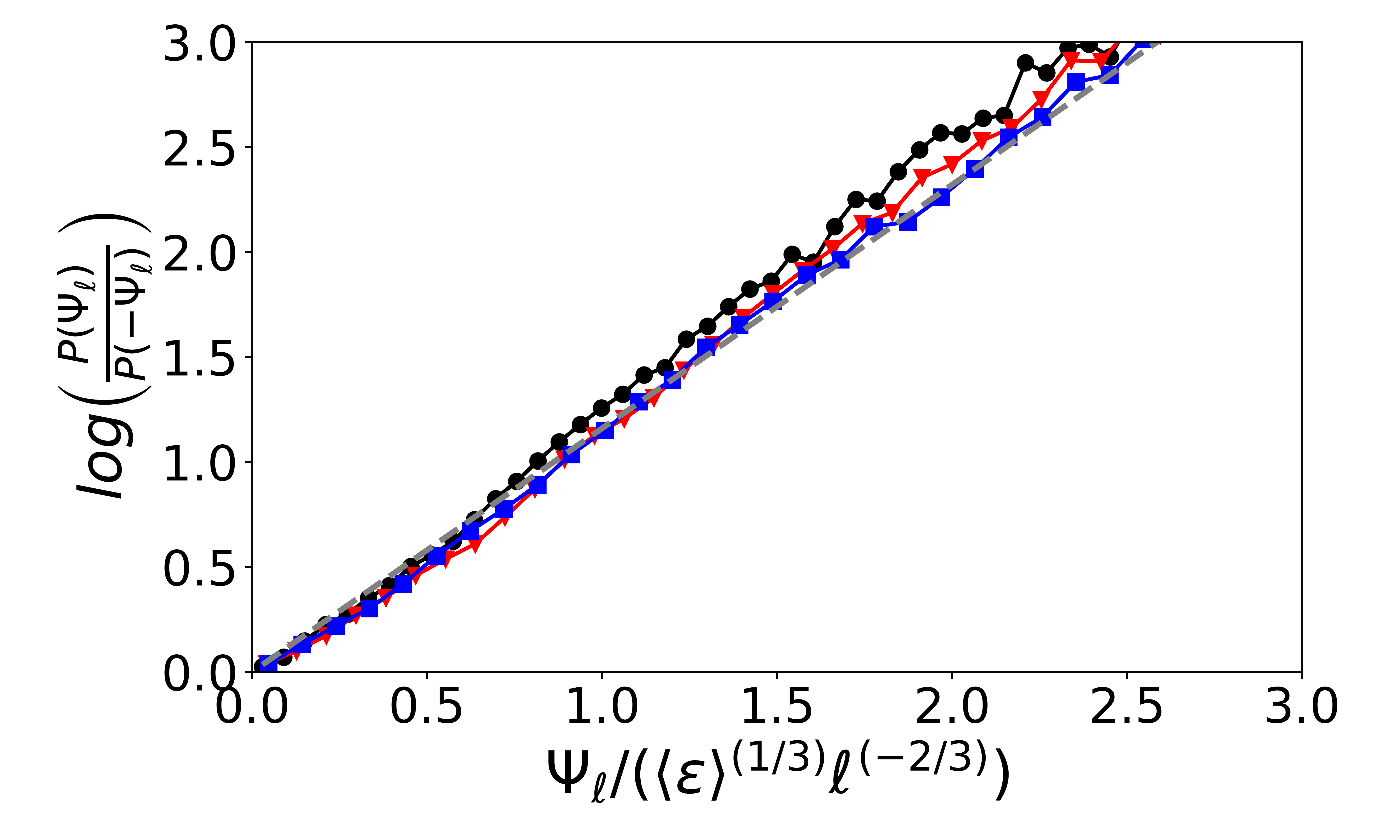}
    \caption{Constant-time Fluctuation Relation test for isotropic turbulence at $R_\lambda = 433$: ratio of probability densities of positive and negative entropy generation rate scales exponentially with the entropy generation rate $\Phi_\ell/k_\ell$ at scale $\ell$. Results are shown for 3 different scales $\ell/\eta = 30$ (black circles), 45 (red triangles) and 60 (blue squares).  The blue dashed line has slope=1.16 obtained via linear fit through the data at $\ell/\eta = 60$.
    }
    \label{fig:FRresult}
\end{figure}

Results are shown in Figure \ref{fig:FRresult}. A close to linear relationship is observed for the logarithm of the probability density ratio of positive and negative entropy generation rate, at the same rate across the three length scales. The dashed line has a slope of 1.16, very close to the earlier result at higher Reynolds number, and confirming that the FR as observed in \cite{yao2023entropy} is valid also at this lower Reynolds number. 

In the following section, we extend our analysis to investigate the FR under finite-time averaging intervals in both Lagrangian and Eulerian time-evolution frameworks. 

\section{Finite time-averaged Fluctuation Relation}
\label{sec:finitetime}

In this section, we examine the finite-time-averaged FR, where the entropy generation rate $\Psi_\ell^\tau$ is evaluated as an average over a time interval of duration $\tau$. This requires specifying how the system evolves over time, which can be approached through either a Lagrangian or an Eulerian framework.

\subsection{Lagrangian framework}

In the Lagrangian framework, the finite-time averaged entropy generation rate $\Psi_\ell^\tau$ is computed along fluid trajectories $\mathbf{\Tilde{X}}(t)$ that move with the sphere of diameter $\ell$. To describe the motion of such a sphere centered at position $\mathbf{\Tilde{X}}(t)$ at time $t$, we use LES-type spatial filtering to evaluate a spherically filtered velocity at ${\bf x}$.  The spatially averaged velocity is denoted as $\mathbf{\Tilde{u}}$, and is obtained by applying a spherical top-hat filter of diameter $\ell$ centered at   position $\mathbf{\Tilde{X}}$,
\begin{equation}
\mathbf{\Tilde{u}} ( \mathbf{\Tilde{X}},t) = \int G_\ell({\bf r}_s) \, \mathbf{u}( \mathbf{\Tilde{X}} + {\bf r}_s,t) \, d^3 {\bf r_s} = \frac{1}{V_\ell} \int \limits_{V_\ell} \mathbf{u}( \mathbf{\Tilde{X}} + {\bf r}_s,t) \, d^3 {\bf r_s}
\end{equation}
where $G_\ell({\bf r})$ represents the filter kernel characterized at length scale $\ell$; ${\bf r_s}={\bf r}/2$ is the radius vector. For the spherical top-hat filter used here $G_\ell({\bf r}_s) = 1/V_\ell$ for $|{\bf r}_s| \leq  \ell/2$ and $G_\ell({\bf r}_s) = 0$  otherwise.  The trajectory $\mathbf{\Tilde{X}}$ is updated at each time step by convecting the sphere center forward using the filtered velocity:
\begin{equation}
    \mathbf{\Tilde{X}}(t + \delta t) = \mathbf{\Tilde{X}}(t) + \mathbf{\Tilde{u}}( \mathbf{\Tilde{X}}(t),t) \, \delta t,
\label{eq:lagrangian_traj}
\end{equation}
where $\delta t$ is the time increment used in the Euler time integration scheme. We use $\delta t$ equal to 5 time-steps of the database, i.e. $\delta t\sim 1/4 \,\, \tau_{K}$ where $ \tau_{K}= 0.0424$ is the Kolmogorov time-scale of the dataset. The choice of $\delta t$ is small enough to ensure accurate resolution of the Lagrangian trajectory and statistics, especially considering that the advection is computed using the filtered velocity field.

This approach effectively captures the time evolution of coarse-grained dynamics of turbulence, providing a physically meaningful time evolution path for evaluating $\Psi_\ell^\tau$.
Moreover, we regard that the sphere of scale $\ell$ moving along this trajectory also as carrying along smaller-scale eddies that are contained inside the sphere of  scale $\ell$. These represent the ``microscopic particles'' of our system. The idea underlying this analysis is that the turbulent entropy being generated is associated with the turbulent fluid in a sphere of diameter $\ell$ that moves with the velocity spatially filtered at that scale. This approach is motivated by the analogous situation in which the standard (thermodynamic) entropy is a material property associated with a small but macroscopic fluid parcel containing many molecules and its temporal evolution is understood in a Lagrangian (material) sense using the local fluid velocity to transport the material. 

\begin{figure} [h]
 \centering
  \includegraphics[scale=0.35]{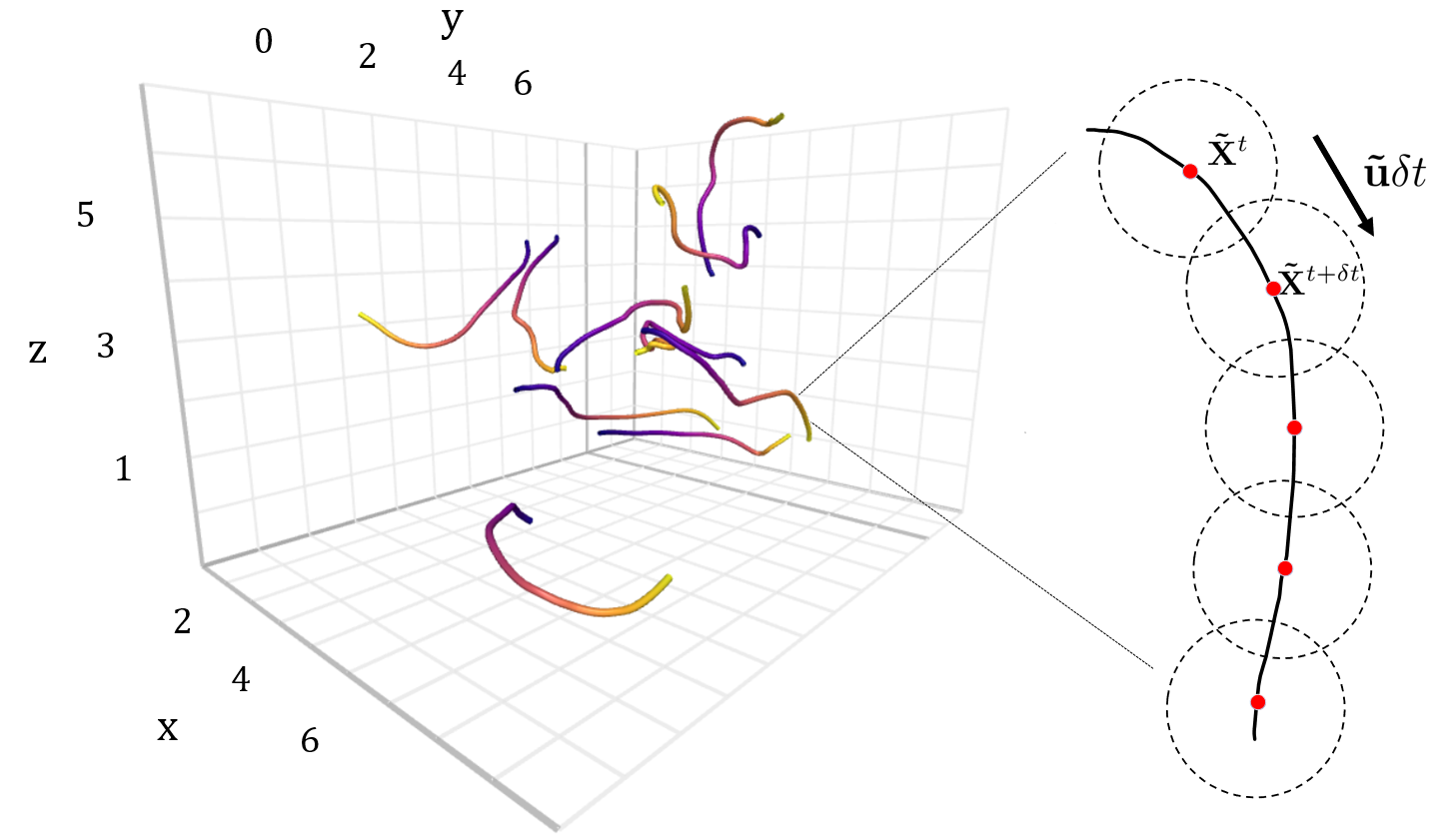}
    \caption{Ten sample trajectories of  spheres (black dashed circles on the right). The diameter of the spheres is $\ell = 45 \eta = 0.126 $ (i.e. very small in the scale of the figure). The convective velocity is taken as the filtered velocity ($\mathbf{\Tilde{u}}$) at the center of each sphere (red dot). Time along the trajectories is indicated by color, ranging from purple (initial time) to yellow (final time), spanning a duration of 500 $\delta t = 2.7 \tau_\ell = 118 \tau_K$.}
    \label{fig:traj}
\end{figure}

Figure \ref{fig:traj} illustrates 10 trajectories moving with the filtered velocity, each initialized from randomly selected coordinates within the $2\pi$ periodic domain. Time along the trajectories is denoted using color, from purple to yellow, representing the time progression from $t = 0$ to $t = 500 \, \delta t = 118 \tau_K$.  

\begin{figure} [h]
 \centering
  \includegraphics[scale=0.35]{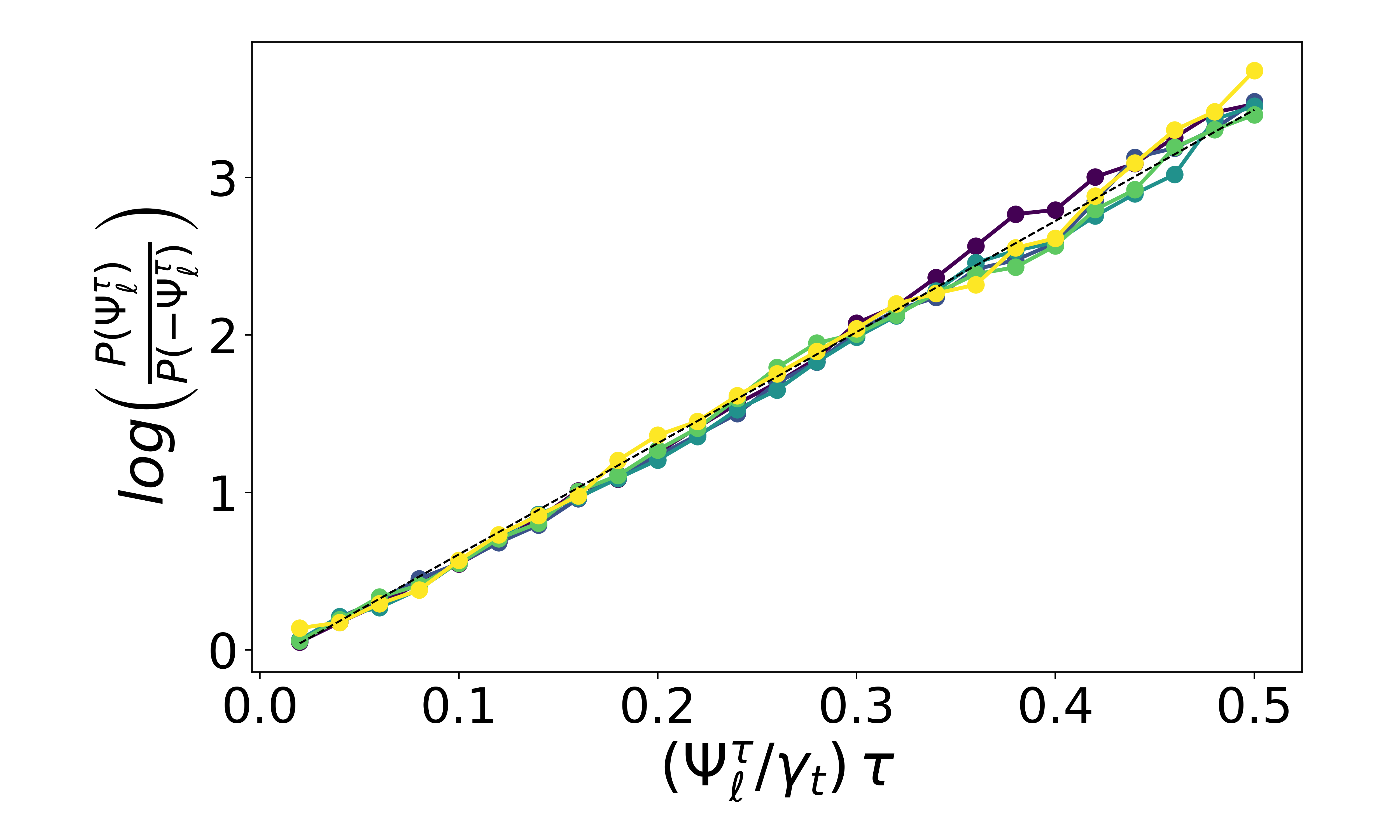}
    \caption{Lagrangian Fluctuation Relation test: ratio of probability densities of positive and negative time-integrated entropy generation  scales exponentially with the time-averaged entropy generation (along Lagrangian trajectories), proportional to $\Phi^{\tau}_\ell/(2$ divided by $2k_\ell$ at scale $\ell = 45 \eta$. Results are shown for 5 different time interval $\tau = 1, 1.5, 2.0, 2.5, 3.0$ (from dark to light color) eddy turn-over time scales, $\tau_\ell = \langle \epsilon\rangle^{-1/3} \ell^{2/3}$. The black dashed line has slope = 7.06 ($= \gamma_t$) obtained via linear fit.}
    \label{fig:Fr_lag}
\end{figure}

The quantities $\Phi_\ell$ and $k_\ell$ are computed at each time along the trajectory, and the time-averaged entropy generation rate $\Psi_\ell/\gamma_t$ (i.e., at first omitting the prefactor $\gamma_t$) is computed by time integration. Specifically we evaluate the finite-time net entropy change
$\Delta_\tau s_\ell \equiv \int_0^\tau (\Phi_\ell/2k_\ell) dt \equiv (\Psi_\tau/\gamma_t) \, \tau$. 
Time integration is performed over 5 discrete integration times $\tau$. We consider time intervals of $\tau = 1, 1.5, 2.0, 2.5, 3.0\tau_\ell$. The results for $\ell = 45 \eta$ are shown in Fig. 
\ref{fig:Fr_lag}. As can be seen, the trend is approximately linear, consistent with the FR. Moreover, the results appear independent of the choice of time-interval $\tau$, implying that the exponential behavior of probability ratios is verified for fluctuations in time-integrated entropy generation over fixed time intervals or also if different time intervals are considered. These observations are consistent with the predictions of the steady-state fluctuation theorem, which asserts validity over arbitrary time durations \citep{seifert2012stochastic}.

Next, we examine the so far unspecified factor $\gamma_t$ that is relevant to our definition of ``temperature''. The ratio of probability densities is independent of the scalar prefactor $\gamma_t$, but the numeric value of $\Psi_\ell^\tau$ is proportional to $\gamma_t$. We obtain an empirical value based on the data by assuming that the FR holds with a slope of unity. The slope in Fig. \ref{fig:Fr_lag} is determined using a least-square error linear fit over all datapoints, with intercept at the origin. The result is $\gamma_t = 7.06$. 

\begin{figure} [h]
 \centering
  \includegraphics[scale=0.35]{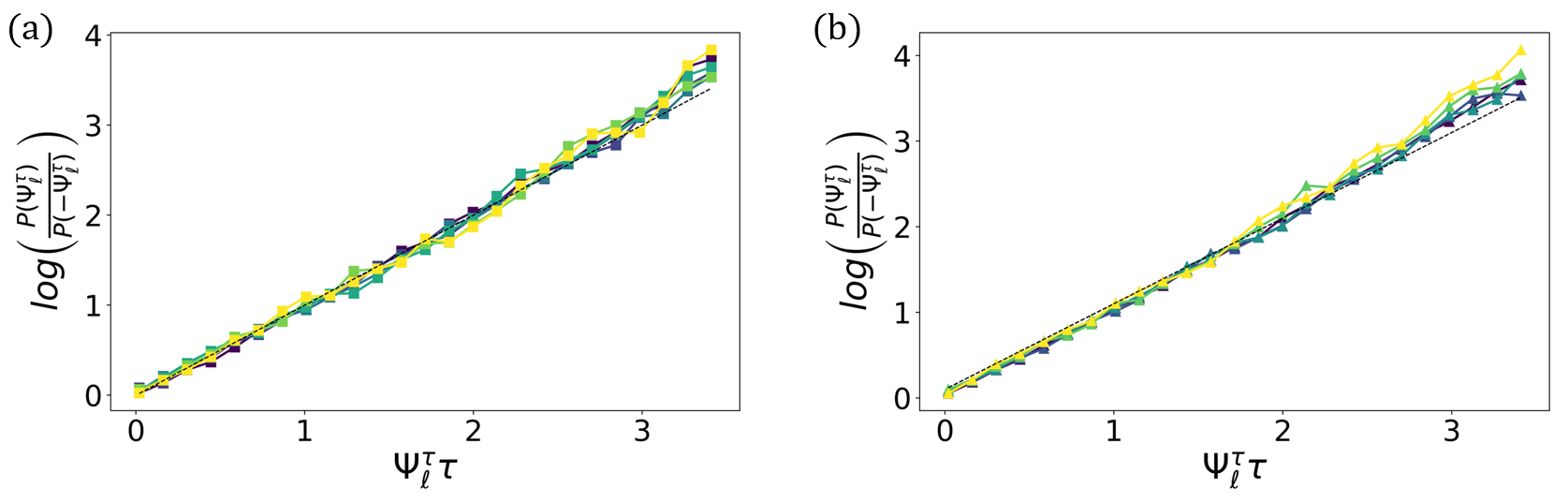}
    \caption{Lagrangian Fluctuation Relation test for three scales $\ell$, each at various time intervals, and using $\gamma_t = 7.06$ in the definition of temperature (and entropy generation rate).   Results are shown for 5 different time interval $\tau = 1, 1.5, 2.0, 2.5, 3.0$ (from dark to light color) eddy turn-over time scales, $\tau_\ell = \langle \epsilon\rangle^{-1/3} \ell^{2/3}$. 
    Squares are for $\ell=30 \eta$ (panel (a)) ;triangles are for $\ell=60 \eta$ (panel (b)). The black dashed line has unit slope. Results conform to a slope of approximately unity, showing that the FR results collapse at various length-scales in the inertial range for the same $\gamma_t=7.06$ value.}
    \label{fig:Fr_lag2}
\end{figure}

We now report results for other values of $\ell$ (namely $\ell=30\eta$ and $60\eta$) and now use $\gamma_t = 7.06$ in the definition of $T$, and therefore $\Psi_\ell$.  Results combining all data for all three length-scales and $\tau$ values are shown in Fig. 
\ref{fig:Fr_lag2}. As can be seen, the FR is valid for other scales as well, also showing linear behavior with a slope or approximately unity when using the same value of $\gamma_t=7.06$. The observations are valid within the range of values we can determine the ratio of PDF's with sufficient statistical convergence ($|\Psi_\ell^\tau \, \tau| \leq 3$). At larger values of $|\Psi_\ell^\tau \, \tau|$, there are too few samples for the negative values and so we cannot determine the PDF on the negative side with sufficient  accuracy.

\begin{figure} [h]
 \centering
  \includegraphics[scale=0.35]{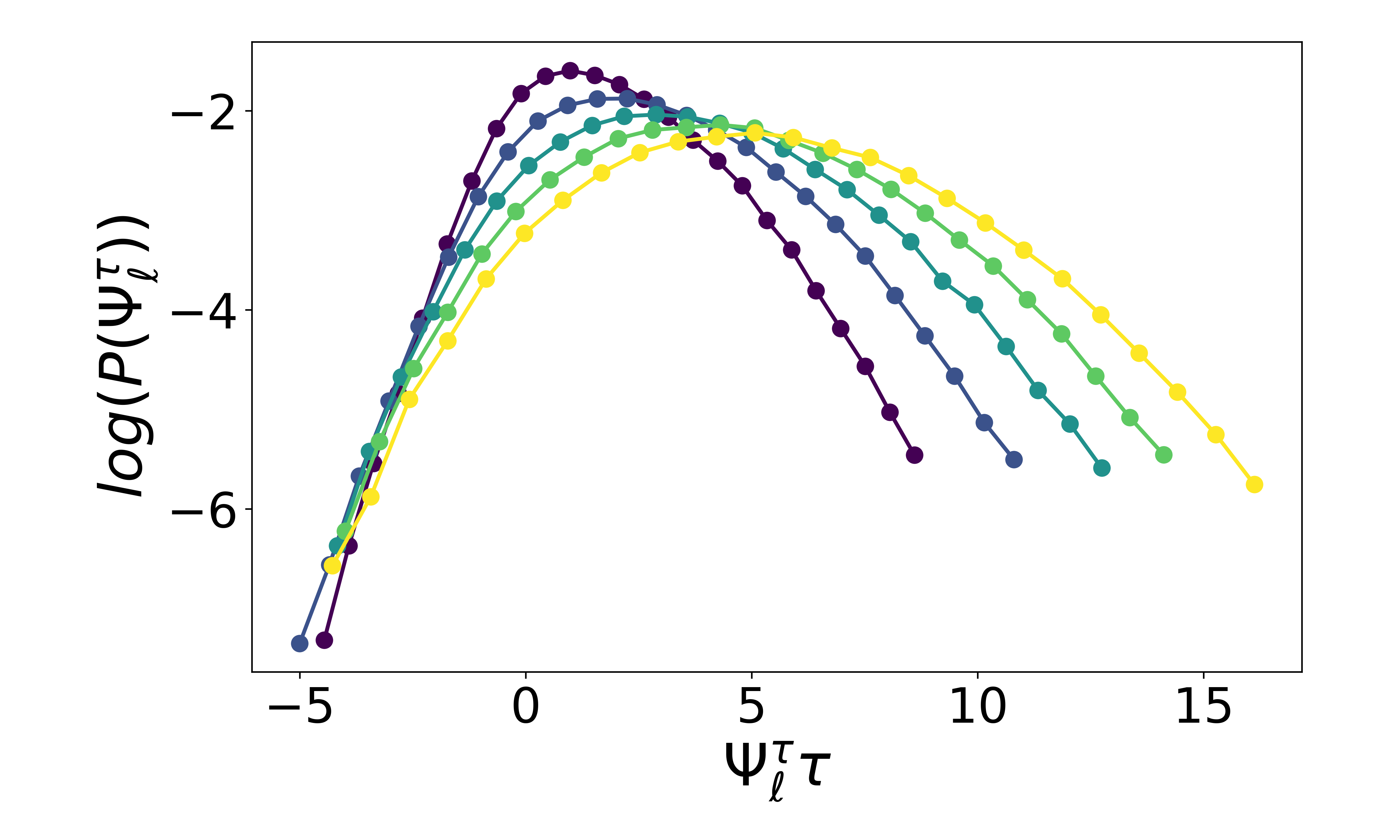}
    \caption{PDFs the finite time-averaged entropy production rate, $\Psi^{\tau}_\ell$, computed along Lagrangian trajectories convecting with the filtered velocity. The results are presented on semi-logarithmic axes for five different time intervals, with color coding consistent with that in Figure \ref{fig:Fr_lag}.}
    \label{fig:PDF_lag}
\end{figure}

Next, we examine the PDFs of the finite time-averaged $\Psi^\tau_\ell$. In previous work \citep{yao2023entropy, yao2024forward}, the PDF of the entropy generation rate at a single time was found to be very closely exponential, on both the positive and negative sides. For such two-sided exponential PDFs, the slope of the linear relationship (in the semi-log plots) demonstrating the FR is simply the difference between the negative and positive slopes \citep{yao2023entropy}. In contrast, for the finite-time-averaged case, we observe a qualitatively different behavior, as shown in Figure \ref{fig:PDF_lag}. For positive values the PDF is curved and close to Gaussian (non-exponential) while at negative values, as the averaging time $\tau$ increases, the PDF gradually shifts from an exponential shape to a more parabolic (or Gaussian-like) profile.  The PDFs have a long, non-Gaussian tails but near the peak, they are more curved, which means that the low-intensity values behave more Gaussian-like than the extreme values at the tails of the PDFs. This transition is consistent with what was reported in several original FR studies for sheared systems, where the PDF of the entropy-like variable becomes Gaussian in the long-time limit (figure 1 of \cite{evans1993probability}). One can attribute this type of trend to the central limit theorem, where increasing $\tau$ effectively aggregates more independent events (e.g., intervals of $\tau$ with non-Gaussian PDFs), leading towards Gaussian behavior when $\tau$ becomes large. Interestingly, regardless of how the PDFs evolve in shape (from one-sided exponential to Gaussian), the FR remains valid. 

\subsection{Eulerian framework}

\begin{figure} [h]
 \centering
  \includegraphics[scale=0.35]{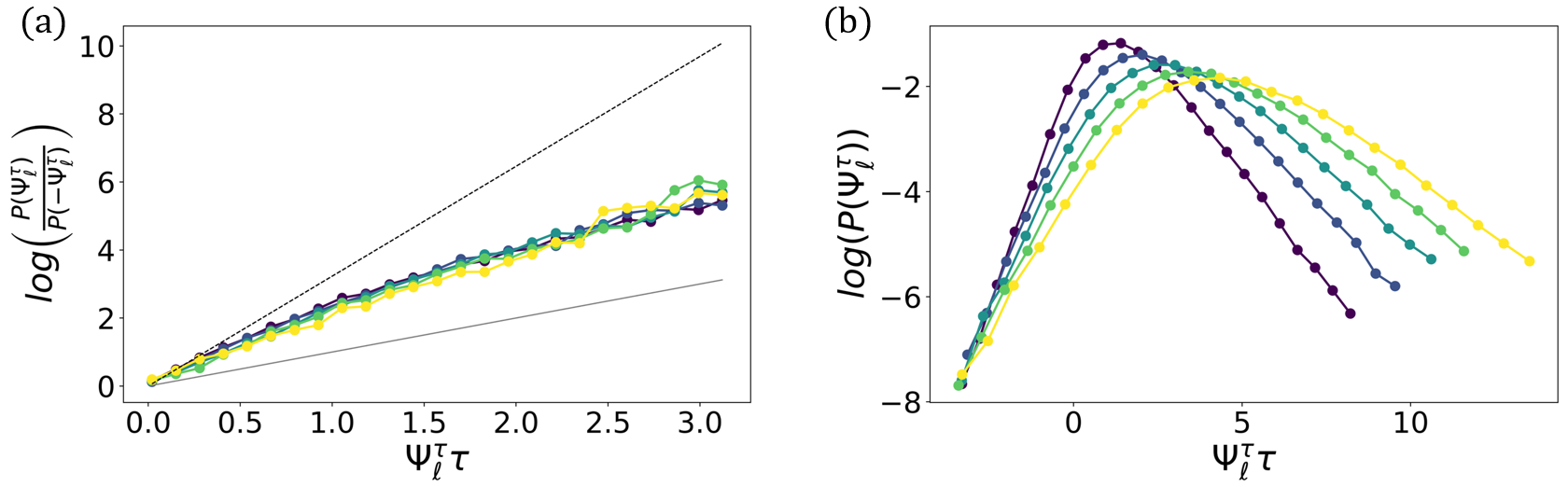}
    \caption{ (a)  Fluctuation Relation test for $\Psi^{\tau}_\ell$ for which time-averaging is computed in an Eulerian fashion, at fixed points; the black dashed line has slope = 3 while the solid gray line has slope = 1.  (b) PDFs of $\Psi^{\tau}_\ell$ computed in Eulerian fashion. Colours represent different time interval $\tau$, with same values and symbols as used in figure \ref{fig:Fr_lag}.
    }
    \label{fig:Fr_Eul}
\end{figure}

In the previous subsection, the time-averaging approach was based on Lagrangian trajectories defined by the filtered velocity field.  To test that the Lagrangian direction represents a special direction along which FR can be reproduced but that other possible directions do not lead to behavior consistent with FR,  we now test the FR in an Eulerian framework. Here the time-averaging for evaluating $\Psi_\ell^\tau$ over a time-interval $\tau$ is performed at fixed spatial locations, effectively setting ${\bf \Tilde{u}} = 0$. In this approach, $\Psi^{\tau}_\ell$ is computed by repeatedly evaluating entropy generation at the same point as the flow evolves past it (still with $\gamma_t=7.06$). In such an Eulerian framework, the entropy generation cannot be assigned or associated with a particular ``turbulent material at scale $\ell$'' (consisting of multiple eddies smaller than $\ell$), but is combining different material and eddies as different fluid elements get advected past the analysis point. In our analysis, the same initial positions and time intervals $\tau = 1, 1.5, 2.0, 2.5, 3.0 \tau_\ell$, and scale $\ell = 45 \eta$ are used, allowing for a direct comparison with the Lagrangian results. As shown in Figure \ref{fig:Fr_Eul}, panel (a), the Eulerian framework yields results that deviate significantly from the  prediction of the FR, for which a linear behavior is expected. Specifically, the FR curve exhibits a  curvature: the curve begins with a slope close to 3 and then at larger values of entropy generation the slope tends to 1. The two slopes are shown as two dashed reference lines with slopes of 1 and 3 across all tested time intervals. This significant deviation from linearity indicates that the FR does not hold in the Eulerian framework.

We also present the PDFs of entropy generation $(\Psi^{\tau}_\ell\,\tau)$ for different (Eulerian) time intervals $\tau$ in Fig. \ref{fig:Fr_Eul}, panel (b). The PDFs in the Eulerian case exhibit a more exponential character, appearing straighter in the logarithmic plot at the tails, and the overall trend between short and long time intervals seems similar to that seen in the Lagrangian analysis. However, the FR significantly deviates from the expected linear behavior since the ratios of probabilities at various values of entropy generation rate do not follow an exponential behavior even approximately.

\begin{figure} [h]
 \centering
  \includegraphics[scale=0.35]{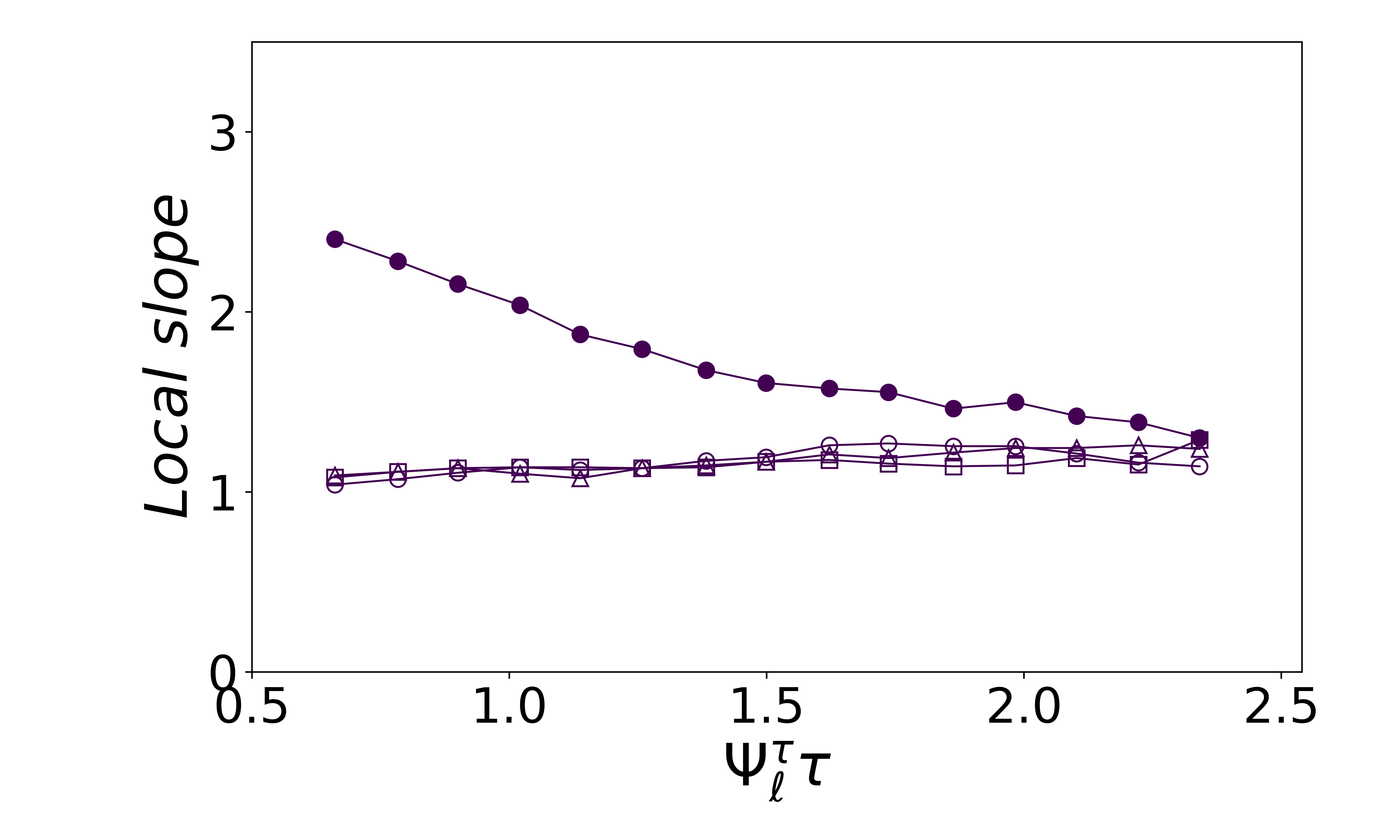}
    \caption{Local slope of relation between $\log(P(\Psi^{\tau}_\ell) /  P(-\Psi^{\tau}_\ell))$
    and $(\Psi^{\tau}_\ell\,\tau)$. The $\Psi^{\tau}_\ell$ is time averaged over one eddy turn-over time ($\tau = \tau_\ell$) at a fixed location (solid symbols) and along Lagrangian trajectories with filtered velocity (open symbols). Square, circle and triangle markers represent results for $\ell = 30, 45, 60 \eta$ respectively.
    }
    \label{fig:Fr_SLOPE}
\end{figure}

To more clearly quantify and compare the degree of linearity in the Lagrangian and Eulerian FR tests (in the semi-logarithmic plot), we compute the local slope of the logarithmic probability ratio as a function of $(\Psi_\ell^\tau \, \tau)$.
Specifically, we perform linear fitting using a sliding set of 10 consecutive points in Figs. \ref{fig:Fr_lag} and \ref{fig:Fr_Eul}(a). The resulting slope is assigned to the mid-point (hence only a range  $0.65 \lesssim (\Psi_\ell^\tau \, \tau)\lesssim 2.35$ is available to display the local slope). The results are shown in Figure \ref{fig:Fr_SLOPE}, where we focus on the case with $\tau = \tau_\ell$. In the Eulerian case (filled black circles), the local slope decreases with increasing $\Psi_\ell^\tau \tau$. This behavior confirms the preceding observation of curvature in the FR plot and increasing asymmetry in the PDFs. By contrast, the Lagrangian framework (open symbols) exhibit an approximately constant slope close to unity, across the entire range of $(\Psi_\ell^\tau \, \tau)$, consistent with the validity of the linear trend expected from the FR in this setting. We also tested the robustness of this result across different inertial-range length scales, specifically $\ell = 30\eta$, and $60\eta$, and found that the constant slope near unity persists in all cases.
(We do take note of a slightly increasing trend in slope for the Lagrangian analysis, but this increase falls below the accuracy with which the slope can be determined, and is in marked contrast to the strongly varying trend of the Eulerian analysis).

The failure of the FR when analyzing data in an Eulerian perspective confirms the importance of considering spatial advection of turbulent fluid parcels. Since the definition of entropy of turbulence is expected to be associated with a material region moving with the flow at scale $\ell$, the observation that the FR holds only when properly following the flow provides support for the definition of entropy of turbulence used in this and prior works \citep{yao2023entropy}.

It is important to note that, even in the Lagrangian framework where trajectories are constructed by convecting with the filtered velocity $\mathbf{\Tilde{u}}$, we are not strictly tracking the same fluid particles over all times.  The group of eddies or particles originally within a given sphere may not remain entirely inside as time proceeds, due to effects such as local shear, turbulent transport across the spherical boundary,  etc.. However, the filtered Lagrangian framework provides a statistically meaningful representation of the advection of coherent structures at the chosen length scale $\ell$ (coarse-grained perspective), capturing the dominant scale-dependent transport dynamics and respecting the causal temporal structure required for evaluating the finite-time FR.

\section{Conclusion}
\label{sec:conclusion}
In this work, we have extended the concept of entropy generation rate in turbulent flows by redefining the original definition proposed in \cite{yao2023entropy}. The key modification involves the definition of the temperature of turbulence, taken to be proportional to the turbulent kinetic energy. Unlike the kinetic theory of gases, where the number of degrees of freedom associated with each microscopic particle is known, in the case of turbulence we do not know this value. Hence it was now left as an empirical parameter $\gamma_t$.  

We first tested the constant-time approximation to the FR using data from forced isotropic turbulence at a moderate Reynolds number, $Re_\lambda = 433$ from JHTDB. We confirmed that the linear trend predicted from the FR holds at this lower-Reynolds number dataset yielding essentially the same results as the higher Reynolds number data considered in \cite{yao2023entropy}. Next, we examined the more fundamental finite time-averaged FR under a Lagrangian framework, where time averaging is performed along trajectories defined by the spatially filtered velocity field. Our results demonstrate that the FR holds robustly, i.e. the results show linear behavior in semi-logarithmic plots of probability ratios versus entropy generation. Various averaging time-intervals were tested, ranging  up to $3\tau_\ell$. In contrast, when the same analysis was performed in an Eulerian framework, the FR no longer holds and the ratio of probabilities no longer follows exponential behavior. This breakdown of the FR might be attributed to the lack of trajectory/temporal coherence, which violates the conditions required for the definition of entropy as relating to fluid material.

While we have demonstrated that the FR holds under filtered Lagrangian trajectories, it remains an open question whether this behavior is unique to the specific choice of trajectory. Future studies should explore alternative definitions, such as trajectories based on unfiltered velocities, filtered velocities in different directions, or flow-aligned coherent structures. Additionally, the connection between physical space trajectories and phase-space trajectories, as described in Fokker–Planck or Liouville frameworks \citep{fuchs2020small}, warrants deeper exploration, particularly in the context of finite-time Lyapunov exponents and attractor dynamics. Finally, it would be natural to extend this study to more complex turbulent systems, including anisotropic turbulence, shear flows, wall-bounded turbulence, and two-dimensional turbulence, where different mechanisms of entropy production and time asymmetry are expected to arise. It will also be of interest to repeat the analysis using Large-Eddy-Simulation (LES) versions of the energy flux and temperature, for which the single-time analysis of \cite{yao2023entropy} showed strong deviations from the FR.  
\vskip6pt

\begin{acknowledgments}
WWe are grateful to the JHTDB/IDIES staff for their assistance with the database and its maintenance. This work is supported by NSF (Grant \# CSSI-2103874).
\end{acknowledgments}

\nocite{*}
\bibliography{aipsamp}

\end{document}